\documentclass[numberedappendix,apj,twocolumn]{emulateapj}
\usepackage[colorlinks=true, citecolor=blue, linkcolor=black,breaklinks]{hyperref}
\usepackage{amsmath,amsthm, amsfonts,amssymb} 

\newcommand{\lya}{Ly$\alpha$}

\newcommand{\HST}{\emph{HST}}

\newcommand{\sigmar}{$\sigma_{r,\mathrm{Ly}\alpha}$}
\newcommand{\fwhm}{$\mathrm{FWHM}_{\mathrm{seeing}}$}
\newcommand{\lyasnr}{$S/N_{\mathrm{int,Ly}\alpha}$}
\newcommand{\zspec}{$z_{\mathrm{spec}}$}

\newcommand{\sigmarall}{$1.70^{+0.09}_{-0.08}$ kpc}

\def\frac#1#2{{\textstyle{{#1}\over {#2}}}}

\def\lsim{\mathrel{\rlap{\lower4pt\hbox{\hskip1pt$\sim$}}
    \raise1pt\hbox{$<$}}}
\def\gsim{\mathrel{\rlap{\lower4pt\hbox{\hskip1pt$\sim$}}
    \raise1pt\hbox{$>$}}}
\def\sqr#1#2{{\vcenter{\vbox{\hrule height.#2pt
         \hbox{\vrule width.#2pt height#1pt \kern#1pt
         \vrule width.#2pt}
         \hrule height.#2pt}}}}

\def\frac#1#2{{\textstyle{{#1}\over {#2}}}}

\def\lsim{\mathrel{\rlap{\lower4pt\hbox{\hskip1pt$\sim$}}
    \raise1pt\hbox{$<$}}}
\def\gsim{\mathrel{\rlap{\lower4pt\hbox{\hskip1pt$\sim$}}
    \raise1pt\hbox{$>$}}}
\def\sqr#1#2{{\vcenter{\vbox{\hrule height.#2pt
         \hbox{\vrule width.#2pt height#1pt \kern#1pt
         \vrule width.#2pt}
         \hrule height.#2pt}}}}

\newcommand{\bea}{\begin{eqnarray}}
\newcommand{\eea}{\end{eqnarray}}
\newcommand{\bit}{\begin{itemize}}
\newcommand{\eit}{\end{itemize}}

\def\picture #1 by #2 (#3){
  \vbox to #2{
    \hrule width #1 height 0pt depth 0pt
    \vfill
    \special{picture #3} 
    }
  }

\def\scaledpicture #1 by #2 (#3 scaled #4){{
  \dimen0=#1 \dimen1=#2
  \divide\dimen0 by 1000 \multiply\dimen0 by #4
  \divide\dimen1 by 1000 \multiply\dimen1 by #4
  \picture \dimen0 by \dimen1 (#3 scaled #4)}
  }

\begin{document}

\title{Constraining Lyman-alpha spatial offsets at $3<z<5.5$ from VANDELS slit spectroscopy}

\author{A. Hoag\altaffilmark{1},
T. Treu\altaffilmark{1},
L. Pentericci\altaffilmark{2},
R. Amorin\altaffilmark{3,4},
M. Bolzonella\altaffilmark{5},
M. Brada\v{c}\altaffilmark{6},
M. Castellano\altaffilmark{2},
F. Cullen\altaffilmark{7},
J. P. U. Fynbo\altaffilmark{8,9},
B. Garilli\altaffilmark{10},
N. Hathi\altaffilmark{11},
A. Henry\altaffilmark{11},
T. Jones\altaffilmark{6},
C. Mason\altaffilmark{12$\dagger$}, 
D. McLeod\altaffilmark{7},
R. McLure\altaffilmark{7},
T. Morishita\altaffilmark{11},
L. Pozzetti\altaffilmark{5},
D. Schaerer\altaffilmark{13,14},
K. B. Schmidt\altaffilmark{15},
M. Talia\altaffilmark{5,16},
R. Thomas\altaffilmark{17}
}

\altaffiltext{1}{Department of Physics and Astronomy, University of California, Los Angeles, CA 90095-1547, USA \email{athoag@astro.ucla.edu}}
\altaffiltext{2}{INAF, Osservatorio Astronomico di Roma, via Frascati 33, 00078, Monteporzio Catone, Italy}
\altaffiltext{3}{Instituto de Investigaci\'on Multidisciplinar en Ciencia y Tecnolog\'ia, Universidad de La Serena, Ra\'ul Bitr\'an 1305, La Serena, Chile}
\altaffiltext{4}{Departamento de F\'isica y Astronom\'ia, Universidad de La Serena, Av. Juan Cisternas 1200 Norte, La Serena, Chile}
\altaffiltext{5}{INAF, Osservatorio di Astrofisica e Scienza dello Spazio di Bologna, via Gobetti 93/3, I-40129 Bologna, Italy}
\altaffiltext{6}{Department of Physics, University of California, Davis, 1 Shields Ave, Davis, CA 95616, USA}
\altaffiltext{7}{Institute for Astronomy, University of Edinburgh, Royal Observatory, Ed- inburgh EH9 3HJ, UK}
\altaffiltext{8}{Cosmic DAWN Center NBI/DTU-Space}
\altaffiltext{9}{Niels Bohr Institute, University of Copenhagen, Juliane Maries Vej 30, 2100 Copenhagen \O, Denmark}
\altaffiltext{10}{INAF, Istituto di Astrofisica Spaziale e Fisica Cosmica Milano, via
Bassini 15, 20133 Milano, Italy}
\altaffiltext{11}{Space Telescope Science Institute, 3700 San Martin Drive, Baltimore, MD 21218, USA}
\altaffiltext{12}{Harvard-Smithsonian Center for Astrophysics, 60 Garden St, Cambridge, MA, 02138, USA}
\altaffiltext{13}{Geneva Observatory, University of Geneva, ch. des Maillettes \altaffiltext{14}{51, 1290 Versoix, Switzerland}
Institut de Recherche en Astrophysique et Planétologie ? IRAP, CNRS, Université de Toulouse, UPS-OMP, 14, Avenue E. Belin,
31400 Toulouse, France}
\altaffiltext{15}{Leibniz-Institut f\"{u}r Astrophysik Potsdam (AIP), An der Sternwarte 16, 14482 Potsdam, Germany}
\altaffiltext{16}{University of Bologna, Department of Physics and Astronomy (DIFA), Via Gobetti 93/2, I-40129, Bologna, Italy}
\altaffiltext{17}{European Southern Observatory, Avenida Alonso de Cordova 3107, Vitacura, 19001 Casilla, Santiago de Chile, Chile}

\altaffiltext{$\dagger$}{Hubble Fellow}

\begin{abstract} 
We constrain the distribution of spatially offset Lyman-alpha emission (Ly$\alpha$) relative to rest-frame ultraviolet emission in $\sim300$ high redshift ($3<z<5.5$) Lyman-break galaxies (LBGs) exhibiting Ly$\alpha$ emission from VANDELS, a VLT/VIMOS slit-spectroscopic survey of the CANDELS Ultra Deep Survey and Chandra Deep Field South fields (${\simeq0.2}~\mathrm{deg}^2$ total). Because slit spectroscopy compresses two-dimensional spatial information into one spatial dimension, we use Bayesian inference to recover the underlying Ly$\alpha$ spatial offset distribution. We model the distribution using a 2D circular Gaussian, defined by a single parameter $\sigma_{r,\mathrm{Ly}\alpha}$, the standard deviation expressed in polar coordinates. Over the entire redshift range of our sample ($3<z<5.5$), we find $\sigma_{r,\mathrm{Ly}\alpha}=1.70^{+0.09}_{-0.08}$ kpc ($68\%$ conf.), corresponding to $\sim0\farcs25$ at $\langle z\rangle=4.5$. We also find that $\sigma_{r,\mathrm{Ly}\alpha}$ decreases significantly with redshift. Because Ly$\alpha$ spatial offsets can cause slit-losses, the decrease in $\sigma_{r,\mathrm{Ly}\alpha}$ with redshift can partially explain the increase in the fraction of Ly$\alpha$ emitters observed in the literature over this same interval, although uncertainties are still too large to reach a strong conclusion. If $\sigma_{r,\mathrm{Ly}\alpha}$ continues to decrease into the reionization epoch, then the decrease in Ly$\alpha$ transmission from galaxies observed during this epoch might require an even higher neutral hydrogen fraction than what is currently inferred. Conversely, if spatial offsets increase with the increasing opacity of the IGM, slit losses may explain some of the drop in Ly$\alpha$ transmission observed at $z>6$. Spatially resolved observations of Ly$\alpha$ and UV continuum at $6<z<8$ are needed to settle the issue.
 
\end{abstract}

\vspace*{0.4truecm}

\section{Introduction}
\label{sec:intro}
The Lyman-alpha (\lya{}; rest-frame 1215.7 \AA{}) emission line has been used as a beacon to spectroscopically confirm the redshifts to the most distant galaxies for decades now \citep[e.g.][]{Steidel+96,Shapley+03,Stark+10,Fink+13,Schenker+14,Zitrin+15b}. It is well-suited for task for two main reasons. First, it is typically the strongest emission line in the rest-frame UV spectra of galaxies. Second, at $z>2$ it is redshifted in the optical/near-infrared where detector quantum efficiency is high, sky backgrounds are relatively low, and it is often the only line accessible for spectroscopic confirmation.

Because \lya{} is detectable out to high redshift and due to its resonance with neutral hydrogen gas, it has been proposed as a diagnostic to probe the state of cosmic reionization \citep[e.g.][]{Haiman+99,Malhotra+04,Fontana+10,Stark+10,Treu+12}. Evidence from the detection of \citet{Gunn+1965} troughs in quasars \citep{Becker+01,Fan+06} suggests that reionization is completed by $z\sim6$, and ongoing at $z>6$. If this is the case, then it is expected that the fraction of Lyman-break galaxies (LBGs) exhibiting strong \lya{} should decline during reionization \citep[e.g.][]{Fontana+10,Stark+10}. This was in fact found to be observationally true \citep{Pentericci+11,Schenker+12,Ono+12,Treu+12,Treu+13,Pentericci+14,Tilvi+14,Schenker+14,Mason+18a,Hoag+19}, providing additional evidence for the onset of reionization at $z>6$. The decline in \lya{} fraction is especially significant for fainter galaxies ($M_{UV}\gtrsim-20.25$) \citep[e.g.][]{Pentericci+14}.

While the drop in \lya{} fraction at $z>6$ is potentially compelling evidence for reionization, especially given the complementary evidence from quasars, many argue that reionization is not the only possible explanation. For example, evolution in Lyman-continuum escape fraction \citep{Dijkstra+14} or increasing numbers of absorption systems present at the end of reionization \citep{Bolton+13} provide potential alternative explanations. Regardless of the reason(s) for the decline, \citet{Haiman+02,Santos+04,Dijkstra+11,Mesinger+15} found that the velocity offset imparted on \lya{} in the ISM and circum-galactic medium (CGM) strongly affects the transmission of \lya{} through the IGM. For example, \lya{} with large velocity offsets ($\gtrsim200~\mathrm{km\, s^{-1}}$) are less attenuated by the IGM. While this is not necessarily evidence against reionization as an explanation of a declining \lya{}, it suggests that the interpretation is complicated due to ISM and CGM physics. 

A less explored explanation for the drop in \lya{} fraction at $z>6$ is differential slit-losses from spatial variations in the distribution of \lya{} emission relative to the UV continuum. This is relevant because most of the \lya{} fraction measurements are made using slit spectroscopy, as this is currently the most efficient way to probe the $z>6$ universe spectroscopically. \lya{} radiative transfer in the ISM and CGM is known to affect both the spectral and spatial distribution of the line. In particular, \citet{Laursen+07,Zheng+11,Dijkstra+12} showed using theoretical models that scattering in the ISM and CGM can produce \lya{} halos an order of magnitude larger in size than the rest-frame UV. Observational evidence of extended \lya{} halos around galaxies first came from narrow-band imaging \citep[e.g.][]{Moller+98,Swinbank+07,Nilsson+09,Finkelstein+11}. The ubiquity of \lya{} halos was later convincingly shown first from stacks of LBGs \citep{Steidel+11} and more recently in individual galaxies \citep{Wisotzki+16,Leclercq+17}. 

Spatially \textit{offset} \lya{} is less well understood. 3D models of \lya{} radiative transfer indicate that \lya{} escape is strongly dependent on the inclination angle \citep[e.g.][]{Laursen+07,Verhamme+12,Behrens+14} in systems with disks. One explanation for this is that \lya{} escapes more readily perpendicular to disks as opposed to through them. This effect might manifest as spatially offset \lya{} along some sight lines. Such offsets have been sporadically reported in the literature. For example, \citet{Bunker+00} demonstrated a convincing \lya{} spatial offset relative to the rest-frame UV continuum in a longslit observation of a bright lensed galaxy at $z=4$. After correcting for lensing, the spatial offset is $\sim1$ kpc. Similarly, \citet{Fynbo+01} found a $\sim4$ kpc spatial offset at $10\sigma$ significance using narrow band imaging. In a more comprehensive narrow band imaging survey, \citet{Shibuya+14} studied a large sample of $z\sim2.2$ \lya{} emitters, finding statistically significant offsets as large as $\sim4$ kpc ($\sim0\farcs5$ at $z=2.2$). The authors do not quantify the frequency or size distribution of such offsets, and they also provide the caveat that some of their offsets are most likely due to mergers. 

The advent of the Multi-Unit Spectroscopic Explorer \citep[MUSE;][]{Bacon+14} on the Very Large Telescope (VLT) has made detailed spatially resolved \lya{} spectroscopy of individual galaxies at $z>3$ possible, without the stringent constraints on the redshift from narrow band imaging. Using MUSE, \citet{Wisotzki+16} found that nearly all of the 26 \lya{} emitting galaxies in their sample have an extended \lya{} halo which is $\gtrsim5-15$ times larger than their rest-frame UV continuum size, and $\sim5$ times larger than \lya{} halos measured in the local universe \citep{Hayes+13,Guaita+15}. However, \citet{Wisotzki+16} were unable to reliably measure spatial offsets between \lya{} and the rest-frame UV due to the large astrometric uncertainty in their HST-MUSE registration, even with an ultra-deep exposure. \citet{Leclercq+17} did observe \lya{} spatial offsets up to $\sim0\farcs3$ in MUSE observations, but they were more interested in constraining the extent of the \lya{} halos. Astrometry issues aside, MUSE has a limited FOV ($\sim 1$ arcmin$^2$), so a statistical measurement of \lya{} spatial offsets would require many MUSE pointings. The MUSE-Wide survey \citep{Urrutia+19} is currently in progress to obtain 100 MUSE pointings at 1 hr per pointing. Assuming the astrometric issues mentioned by \citet{Wisotzki+16} can be overcome, this may be a promising avenue to constrain \lya{} spatial offsets. 

In order to determine the relative importance of the different phenomena that could be causing the drop in \lya{} fraction, it is crucial to first establish the evolution of LBG/\lya{} properties at $z<6$, i.e. after reionization is complete. At redshifts $3<z<6$, \citet{Stark+10,Stark+11,Hayes+11,Curtis-Lake+12,Cassata+15} found that the \lya{} fraction actually increases, possibly due to decreasing fractions of dust and neutral absorbing gas in the inter-stellar medium \citep[ISM; e.g.][]{Finkelstein+12,Jones+13}. A recent study with MUSE, however, found no evidence for an increase in \lya{} fraction over $3<z<6$ \citep{Caruana+18}. The discrepancy in these results may be due to the fact that MUSE integrates the light  over the true \lya{} spatial profile, while slits may miss spatially offset or extended emission. Perhaps the evolution in the \lya{} fraction measured by previous authors with slit-spectroscopy is due to an evolution in the morphology of \lya{} emission which would make it more easily observed in slits at higher redshifts. If this were true, then narrow-band imaging would find a flat \lya{} fraction. \citet{Ouchi+08} investigated the evolution of the \lya{} fraction above an equivalent width of $240$ \AA{} from narrow-band imaging at $z=3.1$, $z=3.7$ and $z=5.7$. The authors found a tentative increase in \lya{} fraction, but the results were statistically consistent with no evolution. 

In this work, we use slit spectroscopy of a large sample ($\sim 300$) of LBGs exhibiting \lya{} in emission to constrain the distribution of \lya{} spatial offsets at high-redshift ($3<z<5.5$). We aim to understand the impact  of slit-losses due to these spatial offsets on the interpretation of current and future \lya{} fraction observations. In Section~\ref{sec:data}, we summarize the data sets that are used in this work. We describe our method for measuring \lya{} spatial offsets in Section~\ref{sec:msr}. In Section~\ref{sec:model}, we describe our Bayesian inference methodology used to recover the physical \lya{} offset distribution. We apply our inference to constrain the offset distribution in Section~\ref{sec:inference} and discuss our findings and their context in Section~\ref{sec:discussion}. We summarize in Section~\ref{sec:summary}. 

We adopt a concordance cosmology with $\Omega_{m} = 0.3$, $\Omega_{\Lambda} = 0.7$,  and $h=0.7$. All magnitudes are reported in the AB system, and all physical distance measurements are in proper kpc, unless otherwise specified.

\section{Data}
\label{sec:data}
Here we describe the spectroscopic and imaging data that we used in this work. 

\subsection{Spectroscopic Data}
\label{sec:specdata}
The primary data used in this work are spectra from the VANDELS survey \citep{Mclure+18,Pentericci+18a}. VANDELS is a deep optical spectroscopic survey with the VIMOS spectrograph on the VLT. VANDELS targeted two fields: one centered on the UKIDSS Ultra Deep Survey (Almaini et al. in preparation; UDS: 02:17:38, -05:11:55) and the other centered on the Chandra Deep Field South (CDFS: 03:32:30, -27:48:28). Both fields have high quality ancillary multi-wavelength data. All spectra were obtained with the VIMOS medium resolution grating. Details on the overview of the survey strategy and target selection can be found in \citet{Mclure+18} while observations and data reduction can be found in \citet{Pentericci+18a}.

The VANDELS team produces 2D (1 spatial axis, 1 dispersion axis) and 1D spectra (dispersion axis only) along with a catalog of spectroscopic redshifts (\zspec{}) of varying quality. The team determined redshifts via the {\tt EZ} software\footnote{\url{http://pandora.lambrate.inaf.it/docs/ez/}} \citep{Garilli+00}. The software allows one to look simultaneously at the 1D, 2D sky and S/N spectra, and also at the \HST{} imaging thumbnails. In the case of \lya{} emission that is spatially offset from the UV continuum, this line will appear in the 2D spectra and will still be identified by the inspectors. Qualities were assigned to the spectra based on the criteria outlined by \citet{LeFevre+05}, using two independent human inspectors. In brief, each inspector assigned a quality of 0, 1, 2, 3, 4 or 9, where 0 means redshift was not able to be measured, 1-4 means the confidence in the redshift measurement was $50\%$, $75\%$, $95\%$ or $100\%$, and 9 means the redshift was assigned based on a single emission line \citep{Pentericci+18a}. The two inspectors were required to come to an agreement on the quality of each redshift after independently grading each spectrum. A final double-check on the flag was done independently by the two co-PIs \citep[see][]{Mclure+18}.

The VANDELS team provide spectra of reference point sources on each mask, but no estimate of the seeing. Because the full-depth spectra were often observed on multiple masks from different nights, the seeing for each object must be calculated individually as a combination of the seeing from the different masks. The seeing we used for each target was the median of the seeing calculated on all of the masks on which it was observed. For each individual mask, which often contained multiple reference point sources, we calculated the seeing by taking the median of the seeing calculated from each reference point source spectrum. For each reference source, we fit a 1D Gaussian to the spatial profile of the spectral continuum. We found that the centroid of the continuum varied significantly with wavelength, so we fit 1D Gaussians in 28 bins of 75 pixels ($\sim190$ \AA{}) each, calculated the standard deviation and then took the median of all standard deviations. The bin size was chosen to balance sufficient $S/N$ for centroiding with the ability to measure wavelength dependence. 

\subsection{Imaging data and VANDELS target selection}
The construction of the photometric catalogs used for target selection are described in detail by \citet{Mclure+18}. To briefly summarize, the VANDELS footprint within the UDS and CDFS fields covers both the central areas which have deep HST imaging as well as the wider areas where only shallower ground-based imaging are available. As a result, a distinct photometric catalog is used for each of the four regions: UDS-\HST{}, UDS-GROUND, CDFS-\HST{}, and CDFS-GROUND. The two regions with \HST{} coverage employ the H-band selected catalogs provided by the CANDELS collaboration \citep{Galametz+13, Guo+13}. The VANDELS team produced photometric multi-wavelength catalogs for the UDS-GROUND and CDFS-GROUND regions as there were no publicly available multi-wavelength catalogs for these regions. The CDFS-GROUND images had variable seeing and were PSF-homogenized to a seeing of $1.0$ arcsec FWHM. The UDS-GROUND images had stable seeing so PSF-homogenization was unnecessary. H-band selected catalogs were produced using 2 arcsec diameter circular apertures for photometry. UDS-GROUND (CDFS-GROUND) spans 12 (17) filters from the U-band to the K-band. For full details on the photometry see \citet{Mortlock+17,Mclure+18}.

For target selection, the VANDELS team made use of the photometric redshifts provided by the CANDELS team for the UDS-\HST{} and CDFS-\HST{} regions \citep{Galametz+13,Santini+15}. For the UDS-GROUND and CDFS-GROUND regions, the VANDELS team generated their own photometric redshifts. The photometric redshifts were derived by taking the median best-fit value from 14 different redshift codes using a broad range of SED templates, star-formation histories, metallicities and emission-line prescriptions. The photometric redshifts from each code were tested and validated against previous spectroscopic redshift data sets from the 2 GROUND field regions, e.g. 3D-HST \citep{Brammer+12,Momcheva+16}, UDSz (Almaini et al., in preparation) in UDS and \citet{LeFevre+05,Vanzella+08,Momcheva+16}.

After removing potential stellar sources from the catalogs, the VANDELS team performed spectral energy distribution (SED) fitting on all sources in the four fields using the \citet{BC03} templates with solar metallicity, no nebular emission, exponentially declining star-formation histories and the \citet{Calzetti+00} dust attenuation law. For more details see \citet{Mclure+18}. 

From this photometric sample in the four regions, potential spectroscopic targets were selected by the VANDELS team to be in these main categories \citep{Mclure+18}: 
\begin{itemize}
\item[(i)] Bright star-forming galaxies in the range $2.4\leq z \leq 5.5$
\item[(ii)] LBGs in the range $3.0 \leq z \leq 7.0$
\item[(iii)] Passive galaxies in the range $1.0 \leq z \leq 2.5$
\end{itemize}

Because \lya{} is only observable at $2.99<z_{\mathrm{spec}}<7.38$ with VIMOS, our spectroscopic sample almost entirely consists of objects in categories (i; 38/305) and (ii; 266/305). The 1 remaining object in our sample came from a sample of Herschel detected objects supplied by D. Elbaz, which was provided after the initial object selection.

The apparent H-band magnitude distribution of the sources in our final spectroscopic sample is well described by a Gaussian distribution with $\mu_H=25.2$ mag and $\sigma_H=0.6$ mag. We determined the absolute magnitude of the sources in our final sample using the H-band magnitude and the \zspec{} recorded by the VANDELS team. The absolute magnitude distribution computed in this way is similarly well described by a Gaussian with $\mu_{MUV}=-20.6$ mag and $\sigma_{MUV}=0.6$ mag.

\section{\lya{} spatial offset measurements}
\label{sec:msr}
To constrain the \lya{} spatial offset distribution, we first assembled a catalog of all galaxies showing \lya{} in emission in the VANDELS database\footnote{\url{http://vandels.inaf.it/dr2.html}}. We queried the database for the UDS\_BEST\_SPECTRA and CDFS\_BEST\_SPECTRA, which represent the full-depth co-added spectra from the UDS and CDFS on January 08, 2019. We note that the slit orientation was always the same for each science target when observed on different masks, such that co-adding the masks does not affect the measured \lya{} offset, if present. Care was taken during the reduction stage to ensure mis-centering did not occur when stacking spectra from multiple masks. We filtered the output by: $2.95<z_{\mathrm{spec}}<8$, which resulted in 611 UDS spectra and 658 CDFS spectra. This filtering step ensured that all of the spectra we downloaded had an assigned spectroscopic redshift, \zspec{}. The medium resolution VIMOS wavelength coverage is sensitive to \lya{} at $2.99<z_{\mathrm{spec}}<7.38$, but we include a wider range in our filtering step in case the redshift assignment was slightly incorrect. Including this extra range will not affect our results as explained below.  We do not perform a filter on redshift quality assigned by the VANDELS team because of the case where a large \lya{} spatial offset could have been misinterpreted as coming from another source. As we point out below, the redshift qualities of the spectra in our final sample turn out to be primarily ($\sim95\%$ of them) Q=3 and Q=4, i.e. $\gtrsim95\%$ confidence in the \zspec{}.

Using the $z_{\mathrm{spec}}$ for each of our downloaded spectra, we visually searched the 2D spectrum for \lya{} emission within a ${\sim30}$ \AA{} window of the predicted \lya{} wavelength. We chose a large spectral window to ensure that we did not miss \lya{} that was offset in velocity from the \zspec{} reported in the catalog. Within this spectral window, we searched the entire spatial extent of the spectrum for \lya{} so that we could detect spatially offset \lya{} emission. We also produced a collapsed spatial profile within this window to visually inspect. After inspecting the 2D spectrum and the collapsed spatial line profile, we flagged each spectrum as either having \lya{} emission or not.  We flagged 426 (194 in UDS, 232 in CDFS) targets as having \lya{}. While we were very inclusive in our visual inspection, in the following steps we removed low $S/N$ and spurious features from our selection. 

To obtain the \lya{} spatial centroid, we first found the optimal spectral line center by collapsing the spectra along the spatial axis and fitting the resulting spectrum to a 1D Gaussian. We used the \lya{} wavelength inferred from the catalog \zspec{} as the wavelength prior for this step. We then produced a \lya{} spatial profile by collapsed the 2D spectrum along the spectral axis in a $30$ \AA{} window centered on the optimal line center we found in the previous step. For this step, we used the imaging catalog position as the spatial prior\footnote{This imaging catalog position is saved under the ``HIERARCH PND WIN\_OBJ\_POS'' keyword in the header of the image extension of each downloaded spectrum fits file.}.

\begin{figure*}[htb]
	\centering
	\includegraphics[width=\linewidth]{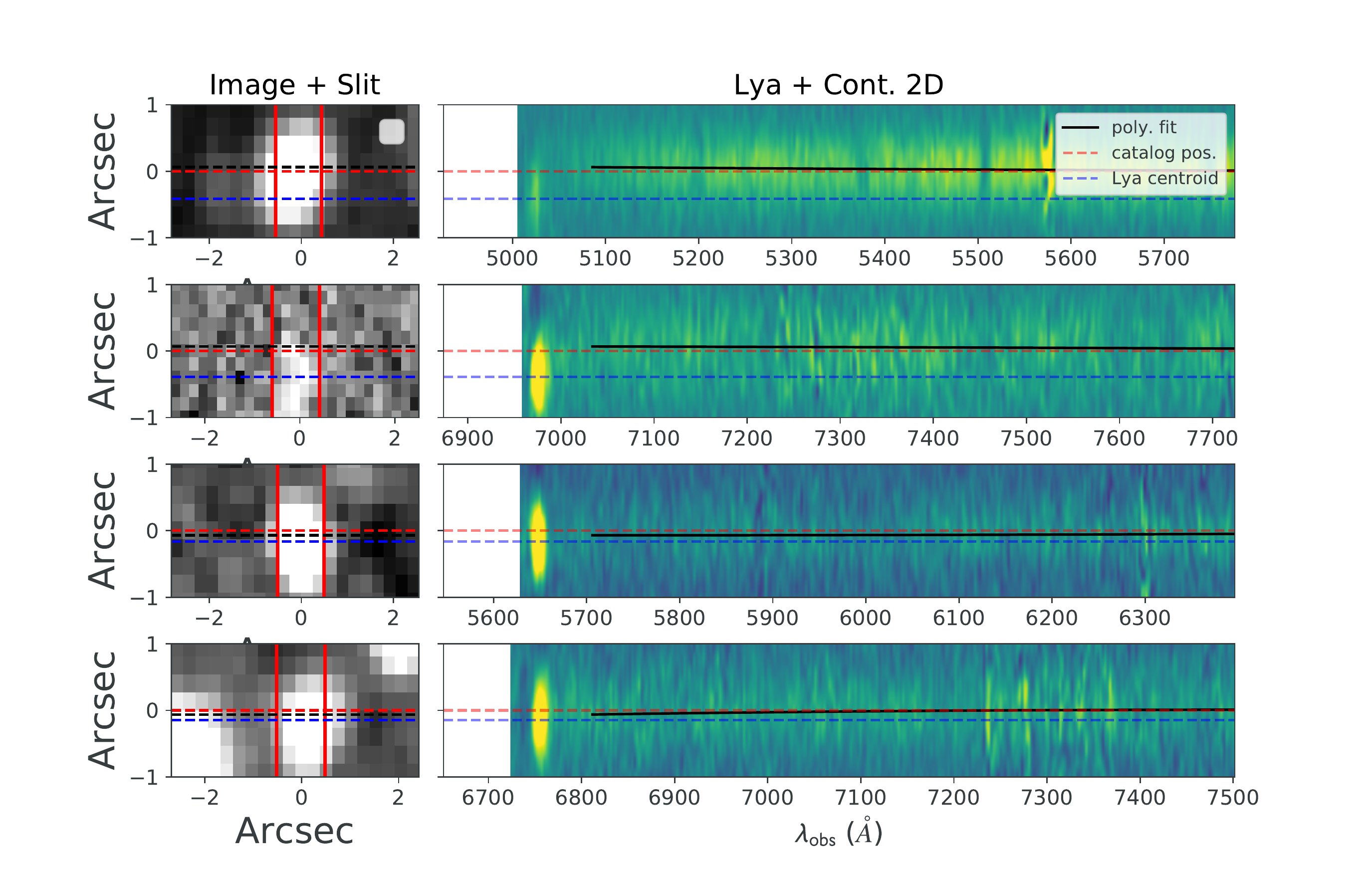} 
	\caption{Illustration of how we compute \lya{} spatial offsets from the VANDELS spectra. {\bf Top row}: A bona-fide spatially offset \lya{}-emitting galaxy. {\bf Left}: r (CDFS) or i (UDS)-band Image of the galaxy showing a portion of the slit overlaid (red vertical lines). Horizontal dotted lines are: (black) continuum position estimated from the polynomial fit (see right panel), (blue) the best-fit \lya{} centroid, and (red) the catalog spatial position which is shown at y=0. {\bf Right}: 2D spectrum from the slit to the left showing the strong \lya{} emission and rest-frame UV continuum, with a clear spatial offset between the two. The continuum emission centroid drifts, so we fit it with a second order polynomial (black line), which we extrapolate to estimate the continuum position at the \lya{} wavelength (dotted black line in left panel). The top two rows are examples of large \lya{}-UV spatial offsets, whereas the bottom two rows show examples of coincident \lya{} and rest-frame UV continuum. }
	\label{fig:offset_examples}
\end{figure*}

While the imaging catalogs provide a spatial continuum centroid in the spectrum, we cannot use this centroid to compare to the \lya{} spatial centroid when calculating the offset between \lya{} and the UV continuum. This is because there is a noticeable drift in the continuum spatial centroid in many of the spectra. Examples of the drift are shown in Figure~\ref{fig:offset_examples}. The drift is likely due to atmospheric refraction, which varies with the airmass of the observations. Because this effect was not corrected for during the reduction, the full-depth reduced spectra, which consist of many sets of exposures taken at varying airmass values, have a blend of spectra with and without the distortion. As a result, in any given spectrum the spectral continuum centroid may be spatially shifted relative to the expected imaging catalog position, and this shift varies with wavelength. 

In order to reliably measure \lya{}-UV spatial offsets in our spectra, we need to be able to calculate the drift. We fit a second order polynomial to the continuum centroid measured in bins of 50 \AA{}, over a bandpass of 1500 \AA{}, starting 35 \AA{} redward of the optimized \lya{} wavelength. We found that $S/N_{1D}\geq2$ per pixel was required for a sufficiently accurate fit to the continuum. We also required an integrated \lya{} signal-to-noise ratio, \lyasnr{}$\geq5$ to accurately measure the \lya{} spatial centroid. We used the 1D signal and noise extensions of the VANDELS data products to calculate $S/N_{1D}$ and \lyasnr{}. These two $S/N$ requirements resulted in 320 objects from the 426 which we flagged as having \lya{}. We visually inspected these objects to assess the polynomial fitting. 15 of the 320 objects had poor fits to their continuum centroid, either due to the presence of continuum from other objects falling serendipitously in the slits or from spectral artifacts. We removed those 15 objects from our sample, resulting in a final spectroscopic sample of 305 objects (131 in UDS and 174 in CDFS).

For the 305 objects in our final sample, we calculated the 1D spatial offset as $y_{\mathrm{Ly}\alpha,\mathrm{opt}} - y_{\mathrm{cont},\mathrm{opt}}$, where $y_{\mathrm{Ly}\alpha,\mathrm{opt}}$ is the mean of the Gaussian fit to the spatial profile of the line and $y_{\mathrm{cont},\mathrm{opt}}$ is the value of the second order polynomial fit to the UV-continuum, evaluated at the best-fit \lya{} wavelength. This polynomial and its extrapolated spatial position are shown for a few example spectra in Figure~\ref{fig:offset_examples}. By using $y_{\mathrm{cont},\mathrm{opt}}$, we correct for the spectral drift, which allows us to reliably calculate an accurate \lya{}-UV spatial offset. 

\begin{figure}[htb]
	\centering
	\includegraphics[width=0.8\linewidth]{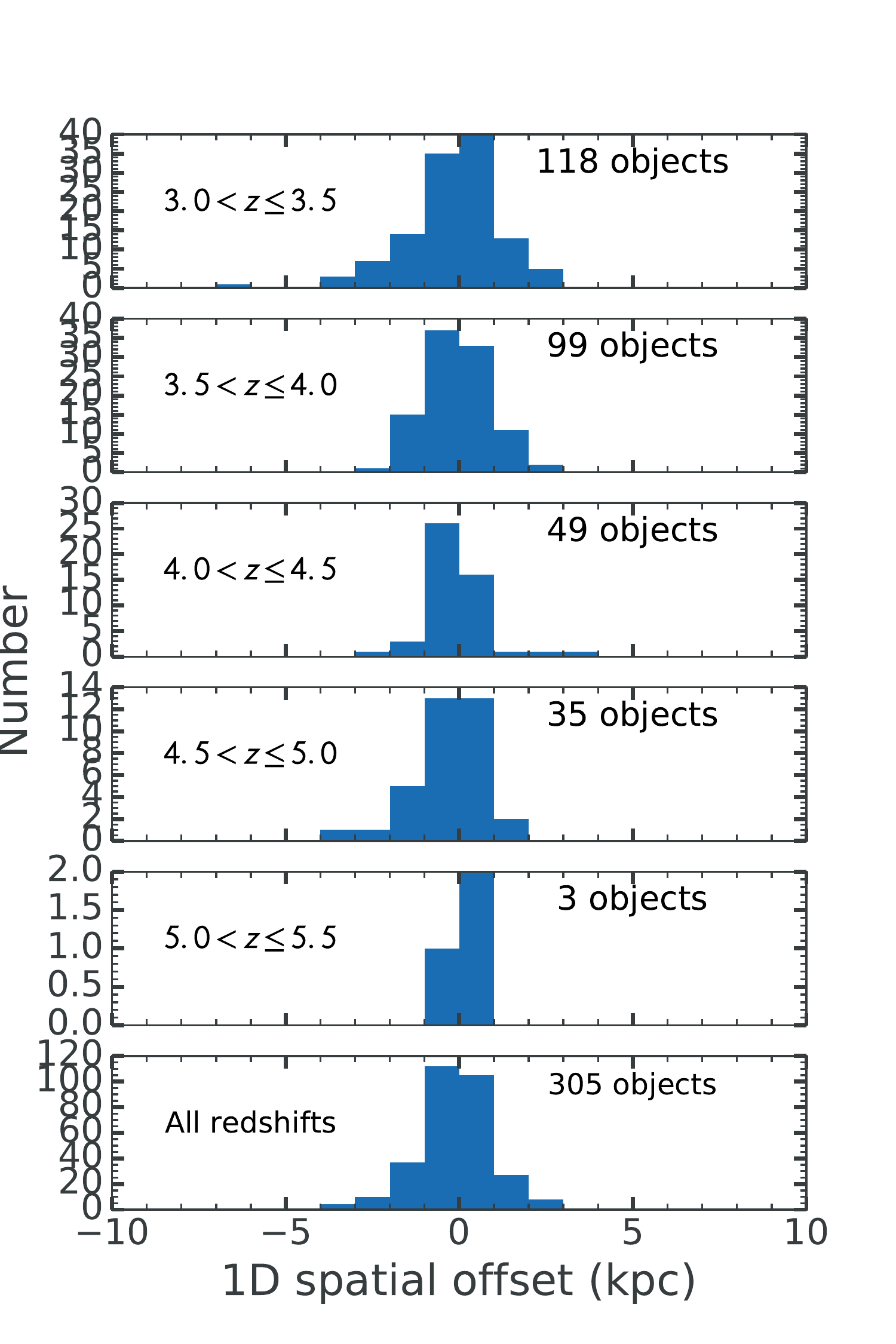} \\
	
	\caption{Distribution of physical 1D \lya{} spatial offsets measured in this work. The top 5 panels show the distributions in distinct redshift bins, while the bottom panel shows the distribution for all objects. Because there is only 1 object in our sample at $z>5.5$ (with offset$=0.3$ kpc), we do not show that bin.}
	\label{fig:offset_hist}
	
\end{figure}

The distribution of 1D spatial offsets measured from the 305 spectra is shown in Figure~\ref{fig:offset_hist}. We note that the maximum spatial offset we can measure in the spectra is set by the slit-length and the position of the target within the slit. VANDELS used variable slit-lengths, but with a minimum of 7 arcsec, or $\gtrsim40$ kpc over the entire redshift range probed in this work, and objects are positioned away from the edges of the slit. Given that the maximum spatial offset that we observe is $<10$ kpc in magnitude, it is extremely unlikely that we missed spatial offsets due to \lya{} falling outside of the slit along the length of the slit. 

Some objects show significantly offset \lya{} emission. For objects with spatial offsets $>2$ pixels ($\simeq0\farcs4 \simeq 2.8$ kpc at $z=4$), we inspected their continuum image thumbnail (from the ``THUMB'' extension of the object fits file) with the slit overlaid to ensure that the emission line was not coming from another object at a different redshift or perhaps an interacting galaxy at the same redshift. The CDFS thumbnails are in the r-band, while the UDS thumbnails are in the i-band. None of these offsets appeared to be coming from a nearby source. We also checked whether the objects with large offsets came primarily from the ground-based imaging part of the VANDELS footprint, but in fact there is an even split (10 in HST coverage, 11 in ground-based coverage) for objects with $>1.5$ pixel offsets. While we cannot completely rule out the possibility that the emission line originates from a source too faint to detect in the continuum image, we stress that it is very unlikely because i) the source would have to have a very high equivalent width (because it is not detected in the continuum image) and ii) the emission line would have to appear at exactly the right wavelength to mimic a spatially offset \lya{}. Such bright serendipitous emission lines with no image counterpart are not detected at other wavelengths in the VANDELS spectra.

In Figure~\ref{fig:offset_examples}, we show two examples of large spatial offsets. The top two rows show cases where the \lya{} emission and UV continuum are likely coming from different regions of the same galaxy. The bottom two rows show cases where the \lya{} emission is spatially coincident with the rest-frame UV continuum. 

\begin{figure}[htb]
	\centering
	\includegraphics[width=\linewidth]{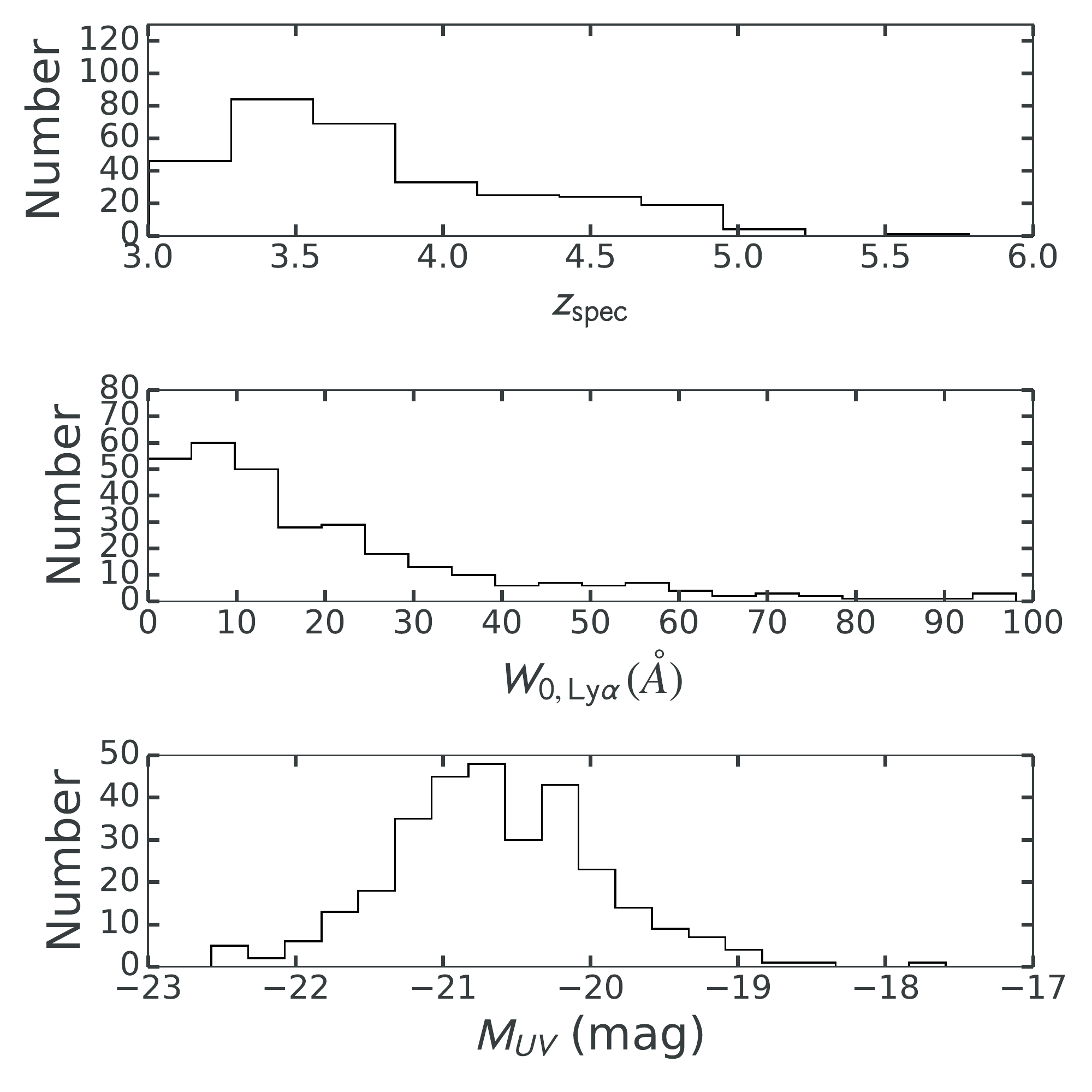} \\
	
	\caption{{\bf Top}: Redshift distribution of the galaxies showing \lya{} emission used to constrain the intrinsic \lya{} spatial offset distribution. The majority of objects have \zspec{}$<4$, and there is only 1 object with \zspec$>5.5$. {\bf Middle}: The \lya{} rest-frame equivalent width ($W_{0,\mathrm{Ly}\alpha}$) distribution of the sample. {\bf Bottom}: The rest-frame UV absolute magnitude distribution of the sample.  }
	\label{fig:redshift_hist}
	
\end{figure}

We show the redshift distribution, \lya{} rest-frame equivalent width ($W_{0,\mathrm{Ly}\alpha}$) distribution, and rest-frame UV absolute magnitude ($M_{UV}$) distribution of the sample in Figure~\ref{fig:redshift_hist}. All but one galaxy in our sample is at $z<5.5$. so this is where we have the statistical power to constrain the \lya{} spatial offset distribution. The equivalent widths were calculated using the \lya{} flux and UV continuum in independent spatial apertures each centered on their respective peak emission, as to account for the potential \lya{} spatial offsets as well as the centroid drift discussed above. To obtain the \lya{} flux, we fit a Gaussian to the 1D \lya{} flux density and sum the Gaussian. A bandpass of 150 \AA{} was used to calculate the continuum flux density from the 1D spectra.

We also investigated the distribution of redshift quality flags of our final sample. There were 0 (Q=0), 4 (Q=1), 4 (Q=2), 86 (Q=3), 203 (Q=4), and 8 (Q=9) spectra with the various quality flags. The vast majority ($\sim95\%$) of our sample has quality flags $Q\geq3$, i.e. confidence of $\geq95\%$ in the \zspec{} assigned to the galaxy by the visual inspectors. We also note that for the 8 ($\sim3\%$) spectra with $Q=9$ (i.e. a definite single emission line), the photometric redshift derived from the deep CANDELS imaging was also used to infer the redshift based on the single line. As a result, we expect that the contamination fraction in our final spectroscopic sample, i.e. the fraction of galaxies that are at a different \zspec{} than the one listed in the catalog, is very small.

The spatial offsets calculated from the 2D spectra are in units of pixels. However, we wish to constrain the physical \lya{} spatial offset, so we convert the pixel offsets to proper $\mathrm{kpc}$, using the \zspec{} to do the angular diameter distance correction. While the redshift inferred from \lya{} may be shifted by up to $\sim500~\mathrm{km/s}$ from systemic \citep[e.g.][]{Shapley+03,Song+14,Verhamme+18}, the difference introduced by this shift into the angular diameter distance is insignificant. We show the distribution of physical 1D \lya{} spatial offsets in proper kpc for the entire sample in Figure~\ref{fig:offset_hist}. The one object in our sample at \zspec{}$>5.5$ has \zspec{}=$5.784$ and a spatial offset of $\sim0.3$ pix $\simeq0\farcs06$ $\simeq0.3$ kpc, which is not a significant offset given the uncertainty of this measurement (see Section~\ref{sec:model}).


\section{\lya{} spatial offset modeling}
\label{sec:model}
Here we describe the Bayesian inference method we employ to constrain the intrinsic distribution of 2D \lya{} spatial offsets from our 1D projected spatial offset measurements described in Section~\ref{sec:msr}. We begin with Bayes' theorem:
\begin{equation} p(m|\{x_i\}) = \frac{p(\{x_i\}|m) p(m)}{E}, \end{equation}
where $p(m|\{x_i\})$ is the posterior for model parameter $m$ given the data set $\{x_i\}$, $p(\{x_i\}|m)$ is the total likelihood, $p(m)$ is the prior on the model parameter, and $E$ is the evidence. 

We choose a simple model to represent the distribution of \lya{} spatial offsets from rest-frame UV center: a circular 2D Gaussian with $\mu_x=\mu_y=0$. This choice is motivated by our ignorance of the shape of the offset distribution with the added bonus that it will allow us to write down the likelihood analytically. The choice to use $\mu_x=\mu_y=0$ is motivated by the fact that there should be no preferred orientation to the offset distribution. Figure~\ref{fig:offset_hist} also provides evidence that the spatial offset distribution is centered at 0. The single parameter we want to constrain is the radial standard deviation of this symmetric Gaussian, \sigmar{}, which we define formally below. We opt not to use the morphology observed in the imaging data as a prior on the \lya{} morphology because we do not want to bias the inference on the \lya{} spatial distribution.

Our data consist of the spatial offset measurements made in the slits. These are 1D projected spatial offsets and therefore do not constitute true two-dimensional offsets. To account for this, we write the Gaussian distribution not in 2D but projected along 1 spatial dimension. 

The 2D likelihood is:
\begin{equation} p(x,y | \sigma_x, \sigma_y) = \frac{1}{2\pi \sigma_x \sigma_y} \exp\left(-\left(\frac{x^2}{2\sigma_x^2} + \frac{y^2}{2\sigma_y^2}\right)\right), \end{equation}
where $\sigma_x$ and $\sigma_y$ are the standard deviations of the Gaussian in the $x$ and $y$ dimensions. For a Gaussian symmetric in $x$ and $y$, $\sigma_x^2 = \sigma_y^2$. And given that $\sigma_r^2 = \sigma_x^2 + \sigma_y^2 = 2\sigma_x^2$, we can write the above more simply as:  
\begin{equation} p(x,y | \sigma_r) = \frac{1}{\pi \sigma_r^2} \exp\left(-\frac{1}{\sigma_r^2}(x^2 + y^2)\right) ,  \label{eq:llh2D} \end{equation}
As we will see, $\sigma_r$ is related to the model parameter we want to constrain, \sigmar{}. The two are not identical because there is measurement uncertainty that we must include in our likelihood function. This will not alter the overall form of the likelihood, so we continue with this expression to derive the 1D likelihood function.

\begin{figure*}[htb]
	\centering
	\includegraphics[width=0.6\linewidth]{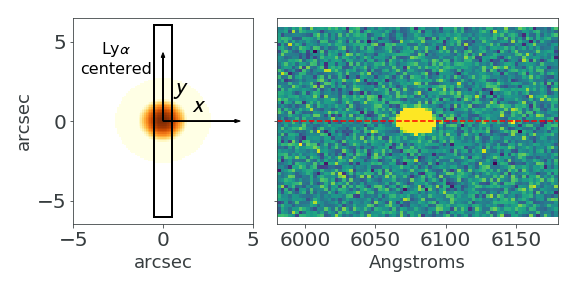} \\
	\includegraphics[width=0.6\linewidth]{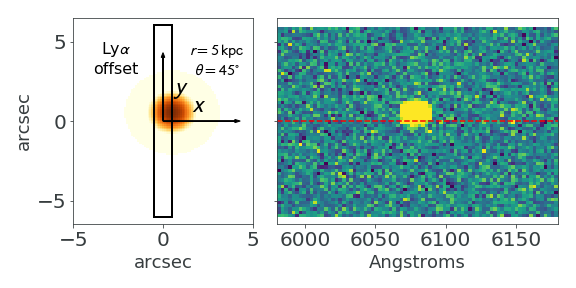}
	
	\caption{Cartoon showing a spectroscopic slit for a simulated Gaussian source at $z=4$ with \fwhm{}$=0\farcs7$. {\bf Top}: Left panel shows the simulated Gaussian source perfectly centered in the slit. Right panel shows a simulated 2D spectrum with the emission line arising from the simulated observation in the left panel. The emission line is not spatially offset in the spectrum. {\bf Bottom}: Same format as top panel, but now the same Gaussian source is offset by $r=5$ kpc ($\sim0\farcs7$) at $\theta=45^{\circ}$ in the left panel. The right panel shows the resulting offset in the spectrum. In this configuration, the $x$-axis is the dispersion axis, so only the $y$-component of the spatial offset in the left panel produces an offset in the spectrum. The $x$-component of the spatial offset results in slit-loss, producing a fainter emission line in the bottom spectrum than in the top spectrum. }
	\label{fig:slitimage}
		
\end{figure*}

When we take a spectrum of a source, we only obtain spatial information along the major axis of the slit, i.e. the axis perpendicular to the dispersion axis. This is illustrated in Figure~\ref{fig:slitimage}. The $y$-axis is the slit major axis, and therefore the $y$-component of the spatial offset is imprinted into the two-dimensional spectrum. As a result, only spatial offsets with a non-zero $y$-component (bottom panel of Figure~\ref{fig:slitimage}) will have a spatial offset in the slit. 


We want to find the likelihood $p(y|$\sigmar{}), where $y$ is the spatial offset we measure in the 2D spectrum. We obtain this by integrating the 2D expression (Equation~\ref{eq:llh2D}) over $x$: 
\begin{equation} p(y|\sigma_r) = \int{dx \,  p(x,y | \sigma_r)} = \frac{1}{\sqrt{\pi} \sigma_r} \exp\left(-\frac{y^2}{\sigma_r^2}\right) \label{eq:llh} \end{equation}
This is simply the equation for a 1D Gaussian centered at $y=0$ with standard deviation $\sigma=\sigma_r/\sqrt{2}$. We note that while the slit width could affect our 2D likelihood, it does not enter the final 1D likelihood $p(y|\sigma_r)$ because it is only a function of $x$. $p(x,y)$ is separable in $x$ and $y$, so no matter what form the slit-width enters into the 2D likelihood, it will always integrate to a constant when doing the $x$ integral in Equation~\ref{eq:llh}.  

If we were able to perfectly measure the spatial offsets in our slit, then we could just substitute \sigmar{} for $\sigma_r$ in Equation~\ref{eq:llh} and then apply Bayes' theorem to obtain the posterior on \sigmar{}. However, there is uncertainty in our measurement of $y$, the projected spatial offset in the slit. This uncertainty depends on a few factors, but most importantly the seeing and the $S/N$ of the \lya{} emission line. If we assume that this uncertainty manifests as Gaussian noise, we can include it in our likelihood via: 
\begin{equation} \sigma_r^2 = \sigma_{r,\mathrm{Ly}\alpha}^2 + \sigma_{r,\mathrm{msr}}^2, \end{equation}
where $\sigma_{r,\mathrm{msr}}$ is the measurement uncertainty on the radial offset. The radial offset uncertainty is related to the uncertainty in the offset in $y$, $\sigma_{y,\mathrm{msr}}$, via $\sigma_{r,\mathrm{msr}} = \sqrt{2} \sigma_{y,\mathrm{msr}}$.

\begin{figure*}[htb]
	\centering
	\includegraphics[width=0.8\linewidth]{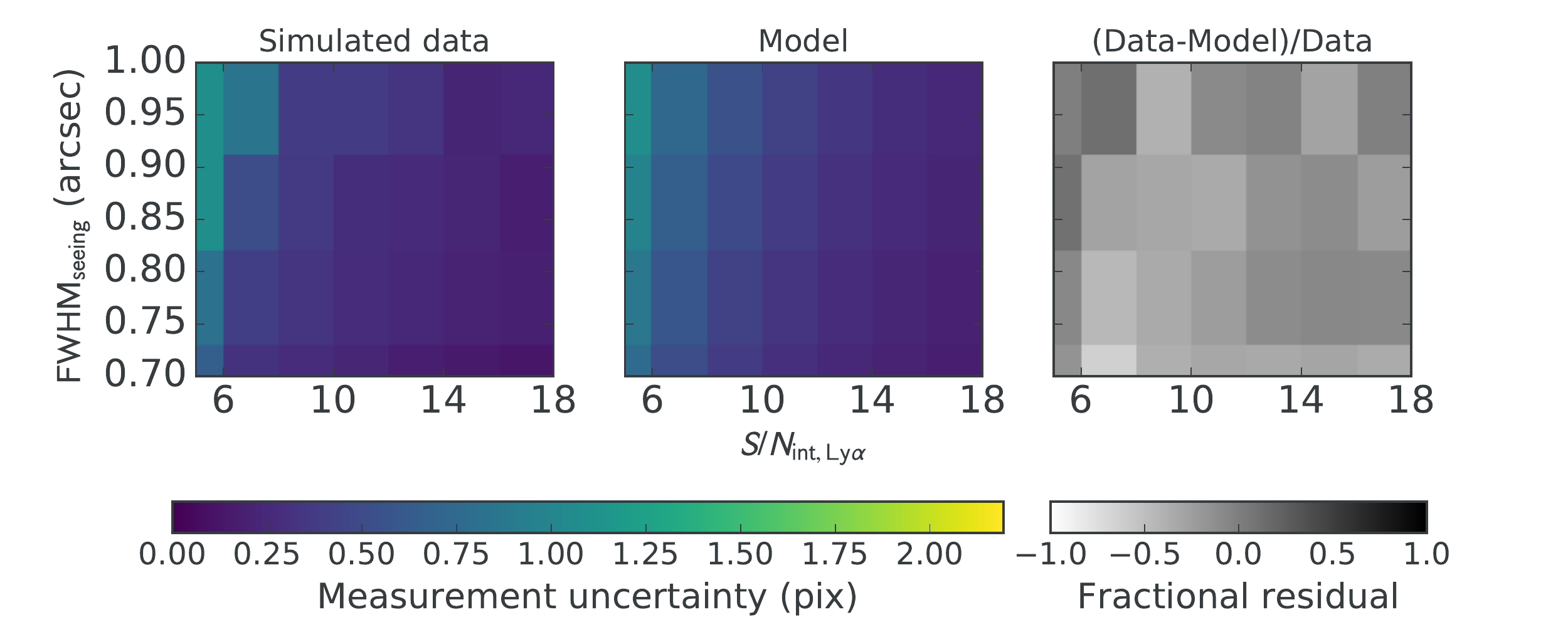}
	\caption{Simulation to estimate the measurement uncertainty on the spatial offset of \lya{} in the 2D spectrum as a function of integrated \lya{} $S/N$ (\lyasnr{}) and \fwhm{}. {\bf Left}: Simulated measurement uncertainty on a grid of \lyasnr{} and \fwhm{}, {\bf Middle}: best-fit model to the simulated measurement uncertainty, {\bf Right}: Fractional residual between the simulated data and model. }
	\label{fig:uncmsr}
	
\end{figure*}

We estimate $\sigma_{y,\mathrm{msr}}$ by performing simulations over a grid of FWHM of the seeing values (\fwhm{}) and integrated \lya{} $S/N$ (\lyasnr{}) values. We use a range of \fwhm{} from 0.1-1.0 arcsec in steps of 0.1 arcsec. For each step in \fwhm, we simulate 1000 \lya{} spectra with \lyasnr{} drawn from the uniform distribution in the range \lyasnr{}$=1-20$, resulting in a total of 10000 simulations. For each simulation, we measure the centroid of the \lya{} emission line using the same process we use to find the line centroid on the real data. We calculate the offset between the correct centroid and the recovered centroid for each simulation. Finally, we take the standard deviation of these offsets in 2D bins given by $\Delta \mathrm{FWHM}_{\mathrm{seeing}} = 0.1$, $\Delta S/N_{\mathrm{int,Ly}\alpha} = 2$ to obtain the estimated measurement uncertainty in each cell. The resulting grid of simulated measurement uncertainties is shown in Figure~\ref{fig:uncmsr}. As expected, the measurement uncertainty decreases with \lyasnr{} and increases with \fwhm{}. 

Given that each spectrum has an arbitrary value of \fwhm{} and \lyasnr{}, we want to be able to estimate the measurement uncertainty based on these parameters. To achieve this, we fit a function to the 2D histogram shown in Figure~\ref{fig:uncmsr}:
\begin{equation} \sigma_{y,\mathrm{msr}} (m,\alpha,c) = m\left(\frac{FWHM}{S/N^\alpha}\right) + c, \end{equation}
finding best-fit values of $m=13.3$, $\alpha=1.4$, $c=0$. This is similar to the standard assumption of $ \sigma_{y,\mathrm{msr}} \sim \frac{FWHM}{S/N}$, but provides a better fit to the simulated data. $\sigma_{y,\mathrm{msr}}$ depends much more strongly on \lyasnr{} than \fwhm{}, especially at low \lyasnr{}. Above \lyasnr{}$=10$, $\sigma_{y,\mathrm{msr}}$ flattens and higher \lyasnr{} do not yield significantly better measurement uncertainty. For the spectra with higher $S/N$ than the range that we simulated, we extrapolate the function to estimate the measurement uncertainty for those objects. For an emission line with the median \fwhm ($0.8$ arcsec) and \lyasnr{} ($10.9$) of our sample, the measurement uncertainty is $\sim0.4$ pixels $\simeq0.08$ arcsec, corresponding to $\sim0.5$ kpc at $z=4$.

With this function in hand, we can rewrite the likelihood for a single spectrum in terms of our model parameter, \sigmar{} and the measurement uncertainty, $\sigma_{r,\mathrm{msr}}$:
\begin{equation} p(y|\sigma_{r,\mathrm{Ly}\alpha}) = \frac{1}{\sqrt{\pi} \sqrt{\sigma_{r,\mathrm{Ly}\alpha}^2 + \sigma_{r,\mathrm{msr}}^2}} \exp\left(-\frac{y^2}{\sigma_{r,\mathrm{Ly}\alpha}^2 + \sigma_{r,\mathrm{msr}}^2}\right)  \label{eq:llh1D} \end{equation}

\section{Inference on the intrinsic \lya{} distribution}
\label{sec:inference}
We wish to find the combined posterior for \sigmar{}, $p(\sigma_{r,\mathrm{Ly}\alpha}|\{y_i\})$, where $\{y_i\}$ are the set of all measured spatial offsets. The combined posterior is simply the product of the individual posteriors, $p(\sigma_{r,\mathrm{Ly}\alpha}|y_i)$. We evaluate the combined posterior using the MCMC sampler from the python package {\tt emcee\footnote{\url{http://dfm.io/emcee/current/}}} \citep{emcee+13}. The inputs to {\tt emcee} are a likelihood, prior, and two parameters specific to the MCMC sampler. We used the likelihood in Equation~\ref{eq:llh1D}  and a flat prior on \sigmar{} over the interval 0-4 kpc. We originally explored a flat prior extending to larger values of \sigmar{}, but the resulting posterior was zero-valued at larger values. The two MCMC parameters are the number of walkers and number of steps per walker. The number of walkers represents the number of independent paths through the parameter space that are taken by the sampler. We use 100 walkers and 250 steps per walker, chosen so that convergence is achieved. We discard the first 30 steps for each walker as these represent the burn-in steps when plotting our posterior or sampling from it to obtain derived quantities.

\begin{figure*}
	\centering
	\includegraphics[width=0.6\linewidth]{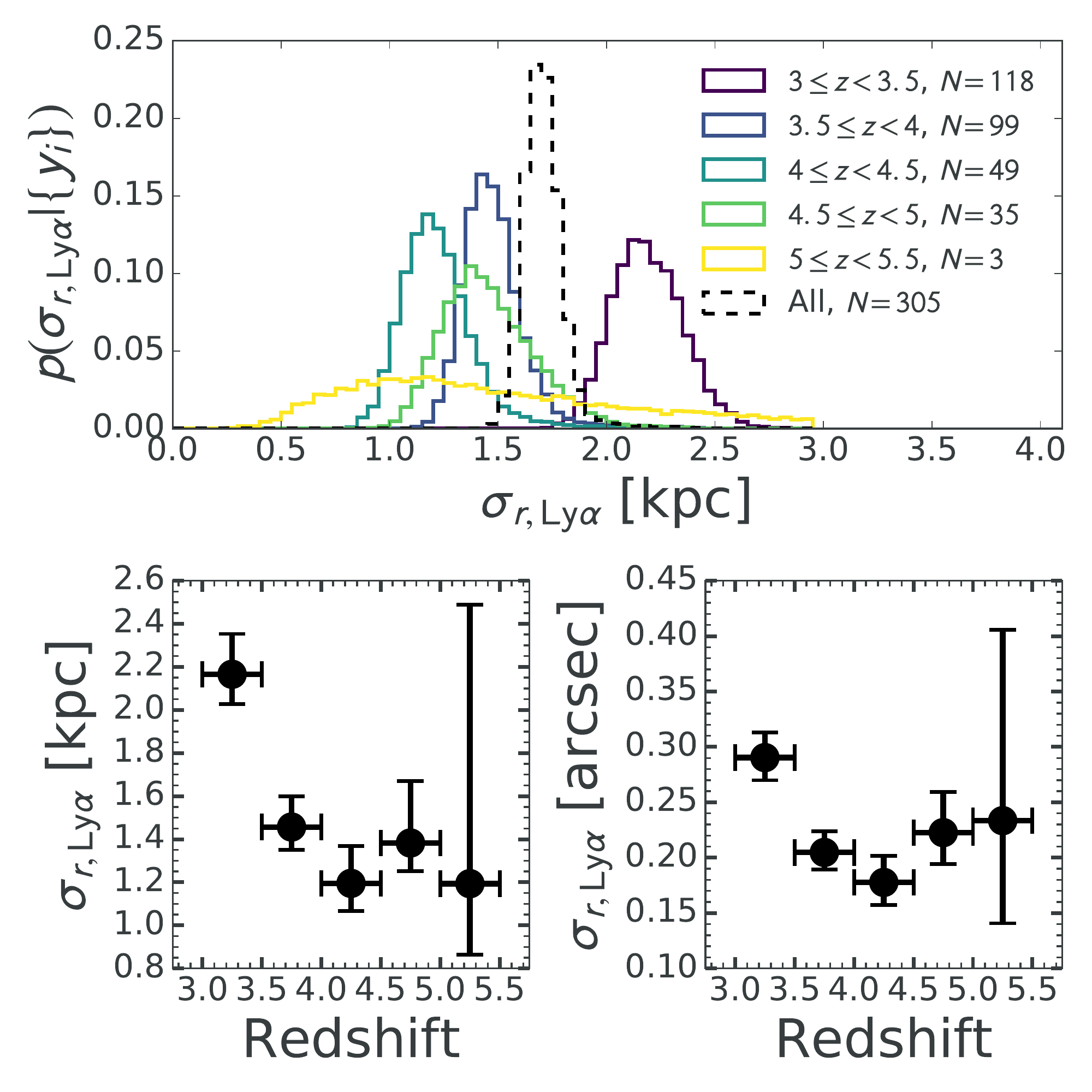} \\
	\caption{{\bf Top}: Posteriors on the standard deviation of the physical 2D \lya{} offset distribution, \sigmar{}, from the VANDELS spectra. We show the posterior for all (black) objects in the sample, as well as posteriors when the data are separated into multiple redshift bins. {\bf Bottom left}: The median and $68\%$ credible intervals on \sigmar{} (kpc) from the posteriors in each redshift bin in the top panel. \sigmar{} shows a decreasing evolution with redshift. {\bf Bottom right}: Same as bottom left but here we show \sigmar{} in arcsec. For reference, the VIMOS slit width is 1 arcsec. In both bottom panels, the horizontal error bars represent the bin width of $\Delta z = 0.5$. }
	\label{fig:posterior}
	
\end{figure*}

We show the final posterior using our entire dataset (305 spectra) in Figure~\ref{fig:posterior}. \sigmar{} is well constrained by our data, with a $68\%$ credible interval of \sigmar{}$=$\sigmarall{}. We also compute the posterior after separating our data into 5 redshift bins: $3\leq z<3.5$, $3.5\leq z<4$, $4\leq z<4.5$, $4.5\leq z<5$ and $5\leq z<5.5$. We show these five posteriors on Figure~\ref{fig:posterior}. Interestingly, the posteriors suggest an evolution to smaller \sigmar{} with increasing redshift. In the bottom two panels of Figure~\ref{fig:posterior}, we show this evolution more clearly. These values, along with the number of objects in each redshift bin, are listed in Table~\ref{tab:conf_tab}. The bottom left panel of Figure~\ref{fig:posterior} shows physical \lya{} offset, whereas the bottom right panel shows apparent \lya{} offset. \sigmar{} declines with redshift over the interval spanned by these data $3<z<4.5$ both in physical and apparent units. At $z>4.5$ uncertainties are too large (due to small numbers) to conclude whether the trend continues. 

To ensure that the decrease in \sigmar{} was not related to the fact that the higher redshift bins have fewer numbers, we re-ran the MCMC in redshift quartiles, i.e. four bins of increasing redshift, each containing an equal number of objects. We also find a decreasing trend in \sigmar{} in the four increasing quartiles: Q1: $2.36^{+0.23}_{-0.21}$, Q2: $1.63^{+0.17}_{-0.15}$, Q3: $1.39^{+0.16}_{-0.12}$, Q4: $1.35^{+0.15}_{-0.12}$.

\begin{deluxetable}{| ccc |}
				\tablecaption{Constraints on \sigmar{} }
				\tablecolumns{3}
				\tablewidth{0pt}
				\tabletypesize{\tiny}
				\tablehead{Redshift bin & \sigmar{} &  $N_{\mathrm{obj}}$ \\ & (kpc) & }  
\startdata
$3.0\leq z < 3.5$ & $2.17^{+0.19}_{-0.14}$ & 118 \\ 
$3.5\leq z < 4.0$ & $1.46^{+0.14}_{-0.11}$ & 99 \\ 
$4.0\leq z < 4.5$ & $1.20^{+0.17}_{-0.13}$ & 49 \\ 
$4.5\leq z < 5.0$ & $1.38^{+0.29}_{-0.13}$ & 35 \\ 
$5.0\leq z < 5.5$ & $1.19^{+1.29}_{-0.33}$ & 3 \\ 
\hline 
All & $1.70^{+0.09}_{-0.08}$ & 305 \enddata
\label{tab:conf_tab}
\tablecomments{$N_{\mathrm{obj}}$ lists the number of objects in each redshift bin.} 
\end{deluxetable}


\section{Discussion}
\label{sec:discussion}
\subsection{Possible causes for \lya{} spatial offsets}
Our inference on \sigmar{} shows that over the redshift interval $3<z<5.5$ \lya{} emission in galaxies can be spatially offset from the rest-frame UV. We also found that the scale of the physical offsets decreases with redshift, at least up to $z=4.5$. While a rigorous theoretical investigation into the origin of these offsets is beyond the scope of this work, we explore two simple hypotheses using our data. 

If the spatial offsets are mostly due to \lya{} scattering, we might expect larger offsets in systems with more scattering, such as more luminous systems and systems with lower $W_{0,\mathrm{Ly}\alpha}$ \citep[e.g.][]{Shibuya+14}. To test this hypothesis, we divided the sample into a ``faint'' bin and ``bright'' bin using the median absolute magnitude, $M_{UV}=-20.67$ mag, to divide them so the two bins had the same number of objects. We then constrained \sigmar{} in both bins using the MCMC approach described in Section~\ref{sec:inference}. We found $\sigma_{r,\mathrm{Ly}\alpha} (\mathrm{faint}) =1.66^{+0.12}_{-0.10}~\mathrm{kpc}$ and $\sigma_{r,\mathrm{Ly}\alpha} (\mathrm{bright}) =1.76^{+0.13}_{-0.12}~\mathrm{kpc}$. The two results are statistically consistent. This could simply be due to our sample size; a larger sample size would allow for a higher precision test which may indicate that either brighter or fainter galaxies have larger \sigmar{}. We also divided the sample into a low equivalent width bin and a high equivalent width bin (both with equal numbers of objects) and inferred \sigmar{} in both bins. We found that \sigmar{} was \textit{larger} in the low equivalent width bin: $\sigma_{r,\mathrm{Ly}\alpha} (\mathrm{low}~W_{0,\mathrm{Ly}\alpha} ) =1.92^{+0.14}_{-0.12}~\mathrm{kpc}$ compared to the high equivalent width bin: $\sigma_{r,\mathrm{Ly}\alpha} (\mathrm{high}~W_{0,\mathrm{Ly}\alpha}) =1.51^{+0.11}_{-0.10}~\mathrm{kpc}$. The difference is statistically significant at $\sim3.5\sigma$. This is evidence in favor of the scattering hypothesis.

A second potential cause for spatially offset \lya{} is dust-screening. In this scenario, dustier regions in the ISM could preferentially absorb \lya{} photons over non-ionizing UV photons. If this were the case, then we might expect galaxies with more dust in general to exhibit a larger \sigmar{}. Using the visual extinction magnitudes ($A_V$) derived from SED fits to the galaxies in our sample \citep{Mclure+18}, we find the opposite to be true. Splitting our sample into two equal-sized bins of $A_v$, we find: $\sigma_{r,\mathrm{Ly}\alpha,} (A_v<A_{v,\mathrm{med}})=1.89^{+0.22}_{-0.16}~\mathrm{kpc}$ for the less dusty half of the sample and $\sigma_{r,\mathrm{Ly}\alpha,} (A_v>A_{v,\mathrm{med}})=1.38^{+0.16}_{-0.13}~\mathrm{kpc}$ for the dustier half, where $A_{v,\mathrm{med}} = 0.25~\mathrm{mag}$. The difference is statistically significant at $\sim3\sigma$. Because less dusty galaxies tend to have larger \lya{} spatial offsets, the dust-screening hypothesis is unlikely, at least with the interpretation we put forward.

The results of the two tests that we performed indicate that the origin of \lya{} spatial offsets may be in part due to \lya{} scattering. In reality it is probably more complicated than the simple scattering or dust-screening hypotheses we put forward due to the complex structure and kinematics of stars and gas in the ISM. Detailed observations of local starbust galaxies, such as are performed by the Lyman Alpha Reference Sample \citep[LARS;][]{Guaita+15} may yield further insight into the origin of \lya{} spatial offsets.

\subsection{Implications for slit-spectroscopic surveys}
The fact that the \lya{} spatial offsets we observe tend to decline in size with redshift is potentially interesting in the context of slit-spectroscopy of \lya{}. For example, several authors have found an evolving \lya{} fraction as a function of redshift over the redshift range probed in this work \citep{Stark+10,Stark+11,Hayes+11, Schenker+12,Cassata+15}. We consider whether the evolution in the scale of the \lya{} spatial offset distribution we infer could partially account for these trends simply due to differential slit-losses. 

In order to test this, we first must convert the physical offset scales, \sigmar{}, into apparent offsets. We show the evolution of the apparent size of the \lya{} offset scale in the bottom right panel of Figure~\ref{fig:posterior}. Like the physical offsets, the apparent offsets inferred from our data exhibit evolution with redshift. The evolution is similar because the apparent size of objects at fixed physical size is relatively flat over the redshift range in our sample: $1~\mathrm{proper~ kpc} = 0\farcs13~(z=3) = 0\farcs14~(z=4) = 0\farcs16~(z=5)$. 

The magnitudes of the largest apparent \textit{radial} offset scales in our sample are \sigmar{} $\sim0.2-0.3$ arcsec at $z\sim3-4$. As illustrated in Figure~\ref{fig:slitimage}, the $x$-component of the radial offset, $r \cos{\theta}$, is what determines the slit-loss. Here we are assuming the slit-length is much larger than the seeing FWHM, which is a good approximation for our VIMOS observations, which use a minimum slit-length of 7 arcsec. We estimate the distribution of slit-losses in each redshift bin by drawing radial offsets from a Gaussian distribution with a standard deviation given by our inferred \sigmar{} in that bin and drawing theta from a uniform distribution. We show the cumulative distribution function (CDF) of slit-losses for the 5 redshift bins in Figure~\ref{fig:slitloss}. As expected, at the redshifts where we inferred the largest \sigmar{} ($z\sim3-4$), the slit-losses are largest. At $3\leq z < 3.5$, $50\%$ ($20\%$) of \lya{}-emitting galaxies will be observed with slit-losses $\gtrsim0.17$ ($\gtrsim0.24$). We note that there is a floor in the slit-loss at $\sim0.04$ for the VIMOS slit/seeing configuration due to the seeing blurring some flux outside of the slit even when \lya{} is perfectly centered in the slit. 

We also explored the slit-losses for a slit-width of $0\farcs5$ to bracket the slit-widths used in ground-based spectroscopy. Because we model the seeing as a Gaussian, this simply shifts the CDFs to along the horizontal axis, i.e. to higher slit-losses. We discuss the implications for the slit-losses on future surveys with smaller slits in Section~\ref{sec:discussion}. 

In calculating the slit-loss, we assume that before convolution with the seeing, the \lya{} emission is point-like. While \lya{} has been shown to have significant spatial extent, often much larger than the UV continuum \citep{Steidel+11,Wisotzki+16,Leclercq+17}, what matters for the following analysis is the differential slit-losses. In this work, we assume that the spatial extent of \lya{} halos is constant on the redshift range we probe. Future spatially resolved \lya{} surveys with VLT/MUSE, for example, may be able to test this assumption at these redshifts.

To assess the impact of these slit-losses on the evolution of the \lya{} fraction, we consider an intrinsic rest-frame \lya{} equivalent width distribution, $p(W_{0,\mathrm{Ly}\alpha})$ before slit-losses and then calculate the differential fraction of \lya{} emitters we would measure if those emitters suffered the slit-losses we found in each redshift bin. We use the  $p_{z\sim6}(W_{0,\mathrm{Ly}\alpha})$ distribution at $z\sim6$ compiled by \citet{deBarros+17} with the parameterization in terms of absolute magnitude ($M_{UV}$) by \citet[][see their eq. 4]{Mason+18a}. Their compilation is the largest sample with a well defined selection function and homogenous observations available at $z\sim6$, i.e. before \lya{} is attenuated by the IGM neutral hydrogen due to reionization. In each redshift bin, we take the product $p_{z\sim6}(W_{0,\mathrm{Ly}\alpha})\times(1-\mathrm{SL})$, where $\mathrm{SL}$ is the slit-loss distribution. We note that in each redshift bin we assume that the intrinsic EW distribution is $p_{z\sim6}(W_{0,\mathrm{Ly}\alpha})$ and that only the slit-losses are evolving with redshift. The result is an adjusted equivalent width distribution in each redshift bin which has suffered slit-losses. The \lya{} fraction is by definition the integrated probability of the distribution above $W_{0,\mathrm{Ly}\alpha}>X$, where $X$ is a threshold value often chosen to be $25$ or $50$ \AA{} for observational convenience. \citet{Stark+11} used $X=25$ \AA{}, so we adopt this threshold when calculating our \lya{} fractions for the purposes of comparison. We generate errors on our \lya{} fractions by resampling from our \sigmar{} posteriors to generate resampled slit-loss CDFs. We then recalculate the \lya{} fraction for each resampled CDF and the underlying distribution, $p(W_{0,\mathrm{Ly}\alpha})$, allowing us to construct \lya{} fraction probability distributions in each redshift bin.

\begin{figure}
	\centering
	\includegraphics[width=\linewidth]{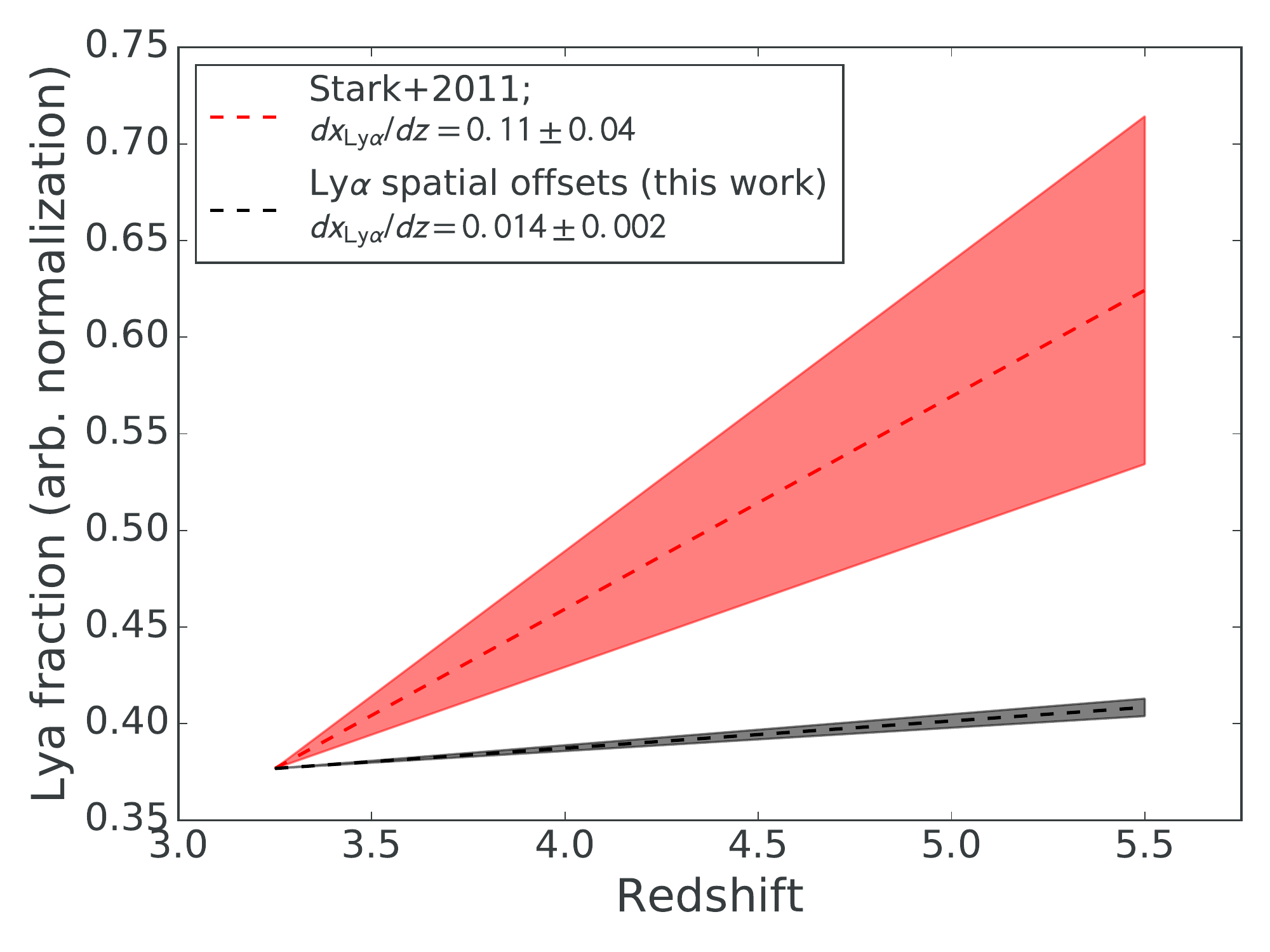} \\
	\caption{Contribution of \lya{} spatial offsets to the differential \lya{} fraction ($dx_{\mathrm{Ly}\alpha}/dz$) over the interval $3<z<5.5$. The red shaded region shows the $68\%$ confidence interval on $dx_{\mathrm{Ly}\alpha}/dz$ from \citet{Stark+11}. The gray shaded region shows the $68\%$ confidence interval on $dx_{\mathrm{Ly}\alpha}/dz$ from the slit-losses induced by \lya{} spatial offsets found in this work. The dotted lines show the best-fit value of $dx_{\mathrm{Ly}\alpha}/dz$ in each case. Both lines are parameterized to go through the same point in our lowest redshift bin, $3 < z < 3.5$, such that the evolution in the \lya{} fraction can be easily compared. The two slopes are inconsistent at $2.4\sigma$, indicating that it is likely that \lya{} spatial offsets are not entirely responsible for the increase in \lya{} fraction observed at $3<z<6$. Given the large uncertainty in the \lya{} fraction evolution, we cannot completely rule out the case where spatial offsets cause the evolution, however.   }
	\label{fig:lyafraction}
	
\end{figure}

We show the differential \lya{} fraction induced by slit-losses that we obtain, $dx_{\mathrm{Ly}\alpha}/dz = 0.014 \pm 0.002$, in Figure~\ref{fig:lyafraction}. This is an order of magnitude smaller than the evolution \citet{Stark+11} found from their sample of $3<z<6$ LBGs: $dx_{\mathrm{Ly}\alpha}/dz = 0.11 \pm 0.04$, although the difference is only $\sim2.4\sigma$. We note that we are only interested in comparing the \textit{differential} \lya{} fraction, $dx_{\mathrm{Ly}\alpha}/dz$, so we scaled the \citet{Stark+11} \lya{} fraction so that it equals the \lya{} fraction measured in our lowest redshift bin. The normalization of the \lya{} fraction is irrelevant in this comparison. \citet{Stark+10} computed $dx_{\mathrm{Ly}\alpha}/dz$ over their fainter luminosity range: $-20.25<M_{UV}<-18.75$, but found that $dx_{\mathrm{Ly}\alpha}/dz$ was similar if they split their sample into two luminosity bins. We adopted $M_{UV}=-20.25$, the mean absolute magnitude from their entire sample, when evaluating the underlying distribution $p(W_{0,\mathrm{Ly}\alpha}|M_{UV})$. We found that if we  instead adopted $M_{UV}=-19.5$, the mean of their faint sample, we obtain a slope of: $dx_{\mathrm{Ly}\alpha}/dz = 0.016 \pm 0.004$, consistent with our fiducial result. We also note that the \citet{Stark+11} \lya{} spectroscopy was performed with the same size slit-widths ($1.0$ arcsec) as the VIMOS slit-widths, which are the slit-widths we assumed in our slit-loss calculations.

The fact that the slit-losses due to \lya{} spatial offsets at $3<z<5.5$ are small is reassuring. Studies inferring the neutral hydrogen fraction during reionization from \lya{} spectroscopy at $z>6$ \citep[e.g.][]{Stark+11,Schenker+14,Treu+12,Treu+13,Tilvi+14,Mesinger+15,Mason+18a,Mason+19,Hoag+19} typically anchor to the \lya{} equivalent width distribution at $z\lesssim6$. If \lya{} spatial offsets produced a large differential evolution in slit-losses at these redshifts, and it was not accounted for, the inferred neutral fractions (and hence the reionization timeline) would be biased. What ultimately matters for these studies is whether slit-losses at $z>6$ are significantly different than what we have measured at $z<6$.  This could arise if the IGM or CGM is patchy on galaxy scales during reionization. UV-bright galaxies may clear channels in their CGM and the IGM through which \lya{} can escape  \citep[e.g.][]{Zitrin+15b,Stark+16,Mason+18b}, potentially resulting in an apparent spatial offset between \lya{} emission and the non-ionizing continuum.  

Finally, we note that in some cases, \lya{} is bright enough to influence the rest-frame UV continuum images. In these cases, the spatial peak of the continuum image may be near the \lya{} even if it is slightly offset. Depending on how common this is, it could mean that \lya{} spatial offsets are already somewhat accounted for in the \lya{} fraction studies. This is easily avoided by using a longer wavelength continuum image whose passband excludes \lya{} during target selection and slit-mask design \cite[e.g.][]{Pentericci+18b,Hoag+19}.

\subsection{Implications for higher redshift \lya{} surveys}
The measurements in this work establish a baseline for \lya{} spatial offsets at $z<6$. The slit-losses we have inferred will impact measured \lya{} fractions at $z<6$ and should be taken into account when \lya{} fractions are reported from slit spectroscopic observations. If spatial offsets measured at $z>6$ are significantly different than our measurements at $z<6$, it can be concluded that the difference is likely due to the neutral IGM during reionization. If the decreasing trend in \lya{} spatial offsets we observed continues out to $z>6$, then \lya{} spatial offsets are not responsible for the decreasing \lya{} fractions noted widely in the literature. In fact, the neutral hydrogen fractions inferred from current studies would need to be higher than reported in order to account for the fact that slit-losses are larger at $z<6$. Given the large uncertainty in \sigmar{} in our highest redshift bin, $5\leq z< 5.5$, and the fact that the trend seems to be changing we cannot meaningfully constrain the impact that anchoring to the $z\sim6$ rest-frame EW distribution would have on reionization studies at $z>6$.

An assumption we made when calculating the differential \lya{} fraction due to slit-losses was that the \emph{spatial extent} of \lya{} halos does not vary over $3<z<5.5$. While there is evidence that \lya{} halos are larger at $z\sim3$ than in the local universe \citep{Wisotzki+16}, it is not clear whether this trend continues to higher redshift. If it does, then slit-losses will become more severe at higher redshift and will result in a steeper slope, $dx_{\mathrm{Ly}\alpha}/dz$, than we measured purely from spatial offsets alone.

While we do not constrain \sigmar{} at $z>5.5$, we can explore what spatial offsets might mean for future spectroscopic surveys at these redshifts. The Near InfraRed Spectrograph (NIRSpec) on JWST will have the capability to perform highly multiplexed multi-object spectroscopy of \lya{} at $z>7$ over a large ($3.6' \times 3.4'$) FOV. However, the effective slit size is $0\farcs2$ (width) by $0\farcs46$ (height) is much smaller than typical slit spectrographs. If \lya{} spatial offsets are not negligible at $z\gtrsim7$, then slit-losses may severely impact the detectability of \lya{} with NIRSpec.

\begin{figure}
	\centering
	\includegraphics[width=\linewidth]{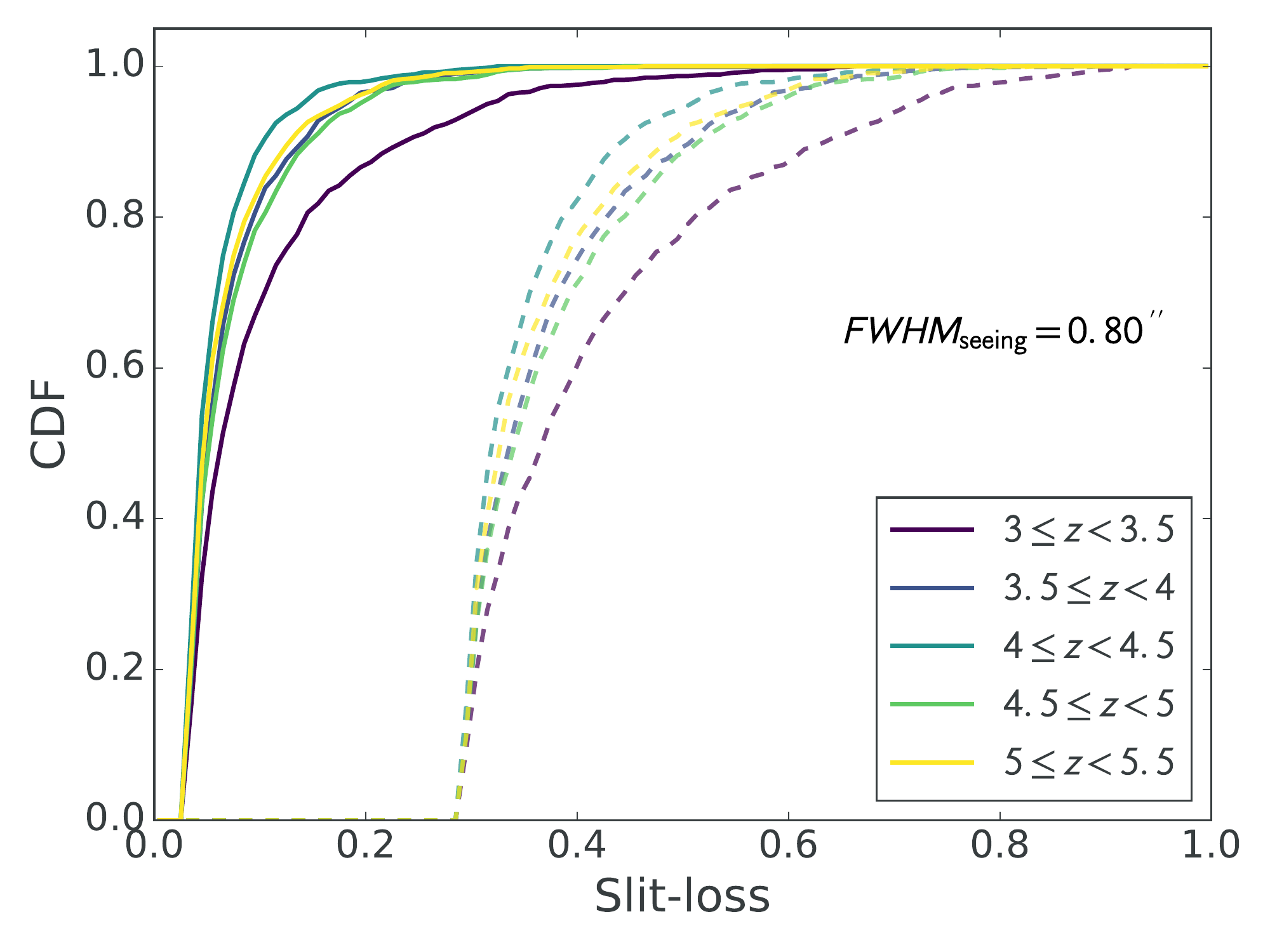} \\
	\caption{Cumulative distribution functions (CDFs) for the slit-loss due to spatial \lya{} offsets. The CDFs are obtained by sampling \lya{} radial offsets from Gaussian distributions with standard deviation determined from our inferred \sigmar{} values and uniform distribution in $\theta$. Solid (dashed) lines show the CDFs when using a slit-width of $1$ ($0.5$) arcsec and a seeing of $0\farcs8$, the median value from our observations. The VIMOS slit-width is $1$ arcsec. The slit-loss is non-zero even for perfectly centered \lya{} due to the seeing.   }
	\label{fig:slitloss}
	
\end{figure}

We forecast the slit-loss distribution at $z=7$ from JWST/NIRSpec observations, using a similar method to the forecast shown in Figure~\ref{fig:slitloss}. We explore two scenarios: i) no evolution in \sigmar{} versus ii) evolution in \sigmar{} relative to $3<z<5.5$. For scenario i), we set \sigmar{} at $z=7$ equal to what we measured over our entire VIMOS dataset ($3<z<5.5$), i.e. \sigmar{}$=$\sigmarall{}. In scenario (ii) we use \sigmar{} at $z=7$ from an extrapolation of a power-law fit to our constraints on \sigmar{} in the five redshift bins. Using this method we find \sigmar{}$=0.93^{+0.01}_{-0.15}$ at $z=7$. In the absence of seeing, the angular size of the emission is governed by the size of the object and its distance. We used the size-luminosity relation compiled by \citep{Kawamata+15} to estimate the UV-continuum size at $z=7$, and then assumed the \lya{} emission is the same size. In reality, \lya{} is generally more spatially extended than the UV \citep[often significantly so up to $z\sim3$;][]{Wisotzki+16}, so our projected slit-losses are likely underestimated. For luminosities comparable to those studied in this work ($M_{UV}\sim-20.5$), this results in effective radii of $\sim0.8 \pm0.2$ proper kpc, or $\sim0.1-0.2$ arcsec at $z=7$. When simulating the slit-loss distributions we draw \lya{} sizes from a Gaussian distribution with $\mu=0.8$ kpc and $\sigma=0.2$ kpc. 

\begin{figure}
	\centering
	\includegraphics[width=\linewidth]{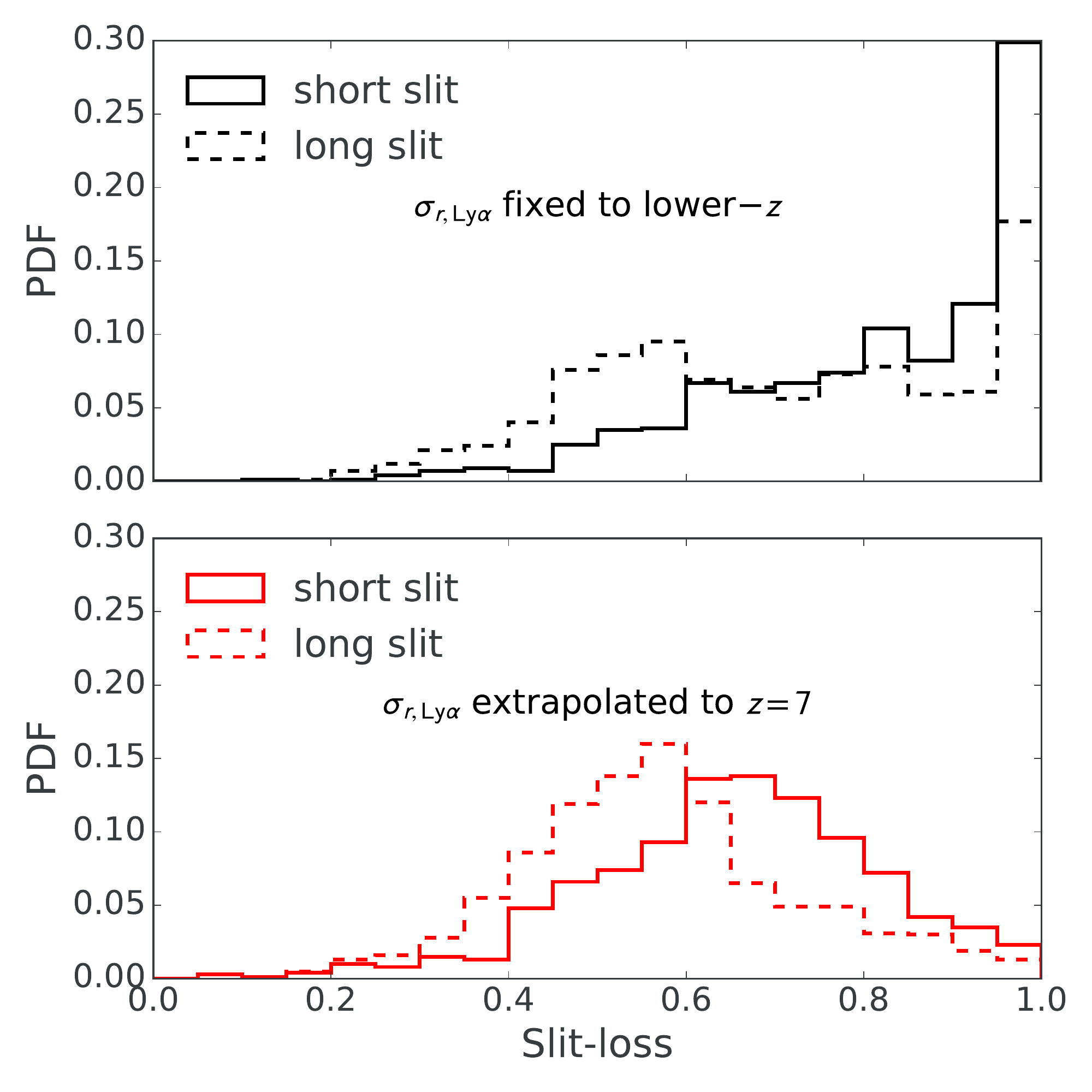} \\
	\caption{Projected slit-losses for JWST/NIRSpec multi-object slit spectroscopy at $z\sim7$ if \lya{} spatial offsets endure to this redshift. The NIRSpec microshutters which form the slits have dimensions $0\farcs2$ wide by $0\farcs46$ tall, much smaller than the VIMOS slits used in this work. We explore two scenarios: {\bf Top:} Expected slit-loss distribution assuming \sigmar{} is the same at $z=7$ as what we measured over our entire dataset ($3<z<5.5$; Table~\ref{tab:conf_tab}). The solid black is for a single microshutter, while the dotted black is for multiple adjacent microshutters which effectively form a longer slit. {\bf Bottom:} Expected slit loss distribution if \sigmar{} is extrapolated from a power-law fit to our binned \sigmar{} measurements. If \sigmar{} does not decrease with redshift out to $z\sim7$, then \lya{} slit-losses with JWST/NIRSpec will be catastrophically large. If \sigmar{} evolves according to our prescription, then slit-losses will still be significant, but less severe. }
	\label{fig:nirspec}
	

\end{figure}

The slit-loss distributions for both scenarios are shown in Figure~\ref{fig:nirspec}. Slit-losses will be significant in either case, but they are largest if \sigmar{} is comparable at $z=7$ to what we measured over the entire redshift range in this work. This is primarily due to the narrow width of the slits. With NIRSpec, one can open adjacent microshutters (assuming the shutter is not disabled), effectively increasing the slit height (cross-dispersion axis). We show how this would affect the slit-loss distributions in both scenarios in Figure~\ref{fig:nirspec}. As expected, it decreases the slit-losses, yet the losses are still significant in either case. While there is a small gap between adjacent microshutters, we ignored these gaps when simulating the slit-losses. As a result, the slit-losses in this scenario are slightly underestimated. Our assumption that the \lya{} size is comparable to the rest-frame UV size likely results in a much larger underestimate of the slit-loss.  

Measuring the spatial distribution of \lya{} emission relative to the rest-frame UV light at $z>6$ will be challenging, in part due to the sensitivity requirements but also because the \lya{} fraction plummets at these redshifts, regardless of the mechanism. Surveys in lensed fields will show \lya{} spatial offsets because spatial offsets in the source plane increase by a factor of $\mu$, the magnification factor, in the image plane. For the same reason, slit spectroscopic surveys in lensed fields will suffer from more severe slit-losses. 

It may already be possible to constrain \sigmar{} at $z\gtrsim6$ with an instrument like Keck/DEIMOS. However, this would likely require a large dedicated effort. The first generation of instruments on upcoming 30-m class telescopes will very likely include wide-field optical spectrographs (i.e. TMT/WFOS, GMT/GMACS, ELT/MOSAIC), which will be capable of tackling this problem. VLT/MUSE also holds promise for constraining \lya{} spatial offsets \citep[e.g.][]{Urrutia+19}, as long as sufficient astrometric precision can be achieved. 

The natural multiplexing of space-based grism spectroscopy is also a potential avenue for constraining \sigmar{}. While the \HST{} WFC3/IR grisms have the wavelength coverage and spatial resolution to detect spatially resolved \lya{} at $z>5.5$, the sensitivity requirements are prohibitive \citep[c.f.][]{Schmidt+16}. With the grisms on JWST/NIRISS and JWST/NIRCAM, it may be possible to constrain \sigmar{} well into the reionization epoch ($z\gtrsim7$).


\section{Summary}
\label{sec:summary}

Using a large sample ($\sim300$) of galaxies showing \lya{} in emission from the VANDELS spectroscopic survey, we constrained the distribution of spatial offsets of \lya{} emission relative to the rest-frame UV continuum. While we used slit spectroscopy which contains less spatial information than, e.g., IFU spectroscopy or narrow-band imaging to constrain the offsets, we employed a large sample which enabled us to make a statistically powerful measurement. 

We parameterized the \lya{} spatial offset distribution with a 2D circular Gaussian with zero mean and a single free parameter, \sigmar{}, the standard deviation of this Gaussian expressed in polar coordinates. We constrained \sigmar{} using Bayesian inference by constructing a likelihood in terms of the measured spatial offset in the 2D VANDELS spectra. Using spectra over the entire redshift range ($3<z<5.5$) in our sample, we inferred a value of \sigmar{}$=$\sigmarall{}. \sigmar{} declines from from $2.17^{+0.19}_{-0.14}$ kpc at $3 \leq z<3.5$ to $1.19^{+1.29}_{-0.33}$ kpc at $5\leq z < 5.5$, or $\sim0\farcs3$ to $\lesssim 0\farcs2$ in terms of apparent size. 

We proposed two simple explanations for the origin of \lya{} spatial offsets: scattering and dust-screening. The fact that \sigmar{} is higher in our lower rest-frame \lya{} equivalent width bin supports the scattering explanation. We found that systems with lower dust content experienced significantly larger spatial offsets, contrary to the dust-screening explanation we put forward. We plan to investigate the origin of spatial offsets further in future work.

We examined the effect that the decreasing spatial offsets would have on slit-losses for \lya{} spectroscopic surveys. Slit-losses alone could result in increasing \lya{} fractions over the range $3<z<5.5$: $dx_{\mathrm{Ly}\alpha}/dz = 0.014 \pm 0.002$. The effect is smaller than, but not entirely inconsistent ($2.4\sigma$) with the \lya{} fraction evolution in the literature, from e.g. \citet{Stark+11}: $dx_{\mathrm{Ly}\alpha}/dz = 0.11 \pm 0.04$. If \lya{} spatial offsets continue to decline with redshift, they are not responsible for the decreased \lya{} transmission measured at $z>6$, typically attributed to reionization. Conversely, if spatial \lya{} offsets become larger as the covering fraction of primeval galaxies increases, then they may represent a significant effect. Future \lya{} surveys with JWST/NIRSpec may experience significant ($>50\%$) slit-losses if \lya{} spatial offsets do not decline more rapidly out to $z=7$ than the redshift evolution in our work suggests. The methodology we developed to infer \sigmar{} at $z<6$ can be readily applied to slit-spectroscopic surveys at $z\gtrsim6$. 

\vspace*{0.5cm}

\acknowledgments{Support for this work was provided by NASA through grant number JWST-ERS-01324 and HST-GO-15212 from the Space Telescope Science Institute, which is operated by AURA, Inc., under NASA contract NAS 5-26555. Support was also provided through NSF grant: COLLABORATIVE RESEARCH: The Final
Frontier: Spectroscopic Probes of Galaxies at the Epoch of Reionization (AST 1815458, AST 1810822). The Cosmic Dawn Center is funded by the DNRF. MB acknowledges support by NSF through award NSF-AAG-1815458 and through NASA ADAP grant 80NSSC18K0945. The authors wish to thank Simon Birrer for helpful discussions during the development of the Bayesian inference analysis.  }


%


  

\end{document}